\documentclass{vgtc}                          

\ifpdf
  \pdfoutput=1\relax                   
  \usepackage{graphicx}                
  \DeclareGraphicsExtensions{.pdf,.png,.jpg,.jpeg} 
\else
  \usepackage{graphicx}                
  \DeclareGraphicsExtensions{.eps}     
\fi%

\graphicspath{{figures/}{pictures/}{images/}{./}} 

\usepackage{microtype}                 
\PassOptionsToPackage{warn}{textcomp}  
\usepackage{textcomp}                  
\usepackage{mathptmx}                  
\usepackage{times}                     
\usepackage{cite}                      
\usepackage{tabu}                      
\usepackage{booktabs}                  

\usepackage{afterpage}

\usepackage{gensymb}
\usepackage{mathptmx}               
\usepackage{enumitem}

\onlineid{4085}

\vgtccategory{Research}

\vgtcinsertpkg

\title{Evaluating 3D User Interaction Techniques on Spatial Working Memory for 3D Scatter Plot Exploration in Immersive Analytics}

\author{Dongyun Han\thanks{e-mail: dongyun.han@usu.edu}\\ %
        \scriptsize Utah State University %
\and Isaac Cho\thanks{e-mail: isaac.cho@usu.edu}\\ %
     {\scriptsize Utah State University }}

\teaser{
  \centering
  \includegraphics[width=\linewidth]{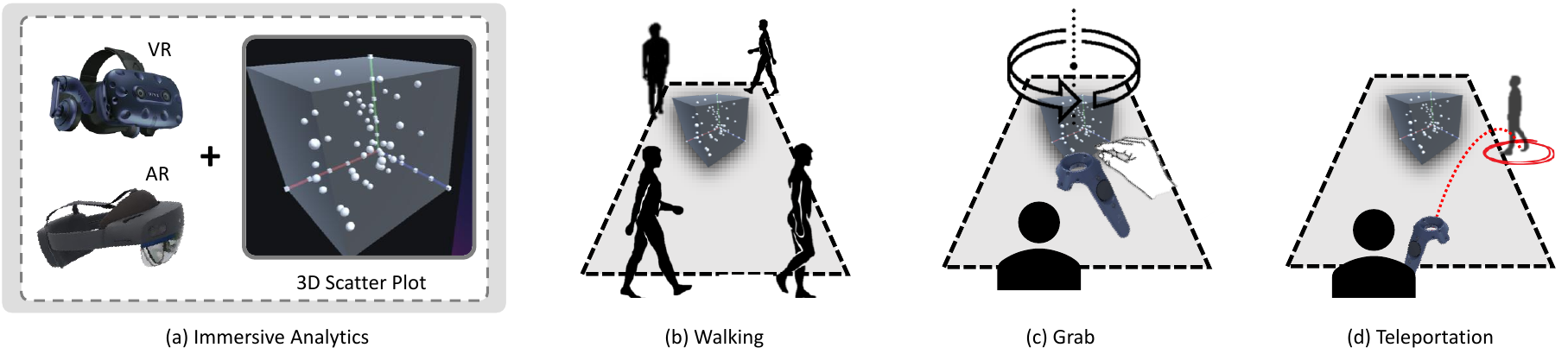}
  \caption{%
  Three 3D user interaction techniques to explore data are evaluated in an immersive analytics setting. (a) Immersive analytics refers to visual analytics performed in an immersive environment using VR and AR technologies. (b) Walking in the real world allows users to navigate immersive environments. (c) Grab allows users to rotate a virtual object by grabbing and turning the object. (d) Teleportation is a widely used navigation technique in VR applications. 
  }
  \label{fig:teaser}
}

\abstract{
This work evaluates three 3D user interaction techniques to investigate their visuo-spatial working memory support for users' data exploration in immersive analytics. 
Two techniques are the common VR locomotion technique, Walking and Teleportation, while the other one is Grab, an object manipulation technique. We present two formal user studies in VR and AR. Our study is designed based on the Corsi block-tapping Task, a psychological test for assessing visuo-spatial working memory. 
Our study results show that Walking supports spatial memory best, and Grab follows. 
Though Teleportation is found to support it the least, participants rated Teleportation as the easiest way to move in the VR study. 
We also compare the Walking and Grab results in the VR and AR studies and discuss differences. 
At last, we discuss our limitations and future work.
} 

\CCScatlist{
  \CCScatTwelve{Human-centered computing}{Visu\-al\-iza\-tion}{Empirical studies in HCI}{};
}

\begin{document}



\firstsection{Introduction}
\maketitle

Over the few years, immersive analytics has attracted many visualization researchers. Immersive analytics indicates carrying out visual analytics in an interactive immersive environment (IE) by utilizing Virtual Reality (VR) and Augmented Reality (AR) technologies. 
Previous studies have suggested that some visualizations could be far more effective in an IE helping users better understand data than in typical 2D desktop environments. For example, 3D scatter plots and node-link diagrams in an IE encourage users to interact with data with a greater sense of immersion, which improves their comprehension of the data and task accuracy~\cite{garcia2016perspectives, butcher2020vria, whitlock2020graphical}. Geographic~\cite{yang2018origin, saenz2017reexamining} and climate visualizations~\cite{helbig2014concept, baltabayev2018virtual} are also studied in immersive settings and evaluated as promising applications. 

Despite growing knowledge of employing immersive analytics, a few research has examined 3D user interaction techniques to support data analytical tasks (e.g., trend/cluster analysis, and data exploring and searching) in IEs.
To provide users with better data analytical tasks in IEs, a deeper understanding of 3D user interaction techniques is essential~\cite{chandler2015immersive, marriott2018immersive}. 
Visualization researchers have compared the usability of these techniques for immersive analytics for data search tasks and a general understanding of 3D plots (e.g., trends and outliers). These are discussed in Section~\ref{sec_relatedWork2}. 

Understanding spatial working memory support of interaction techniques is critical in visual analytics~\cite{larkin1987diagram, lohse1997role, borkin2015beyond}. The ability to search and recall spatial placements of visualization elements within an information space is important to perform visual analytical tasks~\cite{liu2022effects}. 
For example, a data analyst begins his or her task by identifying data of interest within a visualization. The analyst then examines the visualization and seeks new information by comparing the newly found data with previously noted data held in memory. As the data exploration progresses, the analyst can discover trends from the set of data explored or anomalies that deviate from the trend. To perform this series of activities (e.g., data comparison, exploration strategy planning, trend and anomaly detection~\cite{amar2005low, brehmer2013multi}), the analyst relies on memorizing and referencing spatial information of data he or she previously explored. Unlike common 2D visualization with a desktop, data exploration in immersive analytics is accompanied by visualizations or users' movement in a 3D space, and requires higher spatial memory ability. Previous research has shown that physical body activities have an impact on spatial memory ~\cite{friedrich2021effect, radle2013effect}. 
However, it is unclear how 3D user interaction methods impact users' spatial working memory for immersive analytics. 

This paper addresses the gap in knowledge regarding the effectiveness of 3D UI for data exploration for immersive analytics. 
We conduct two formal user studies to evaluate the effectiveness of 3D user interaction techniques to support spatial working memory in VR and AR.
For the studies, we utilized the Corsi block-tapping test~\cite{corsi1972human}, a psychological test assessing an individual's visuo-spatial memory. The first study evaluates three interaction techniques, \textit{Walking}, \textit{Teleportation}, and \textit{Grab} in a VR environment. 
Walking and Teleportation are commonly used navigation techniques in VR. 
Grab is a 3D object manipulation technique to control object orientation. 
The second study compares spatial memory performance with two interaction techniques, \textit{Walking} and \textit{Grab} in AR. Our results of both studies show that Walking supports visuo-spatial memory well in a data exploration task, and Grab follows. 
The VR study shows Teleportation supports the lowest visuo-spatial memory compared to the other two techniques. 
Interestingly, however, participants rated Teleportation as the easiest interaction method. 
We further compare and discuss the VR and AR results. 
Our findings suggest that tapping tasks in AR are more accurate than those in VR, despite the fact that interaction time in VR is much faster. 
At last, we report limitations and future work.



\section{Related Works}

\subsection{Immersive Analytics}\label{sec_relatedWork1}

Immersive analytics removes barriers between analysts and data to explore and analyze data-driven problems in the immersive settings of VR and AR. Earlier studies report that spatial analysis tasks such as distance estimation~\cite{etemadpour2013effect}, cluster identification~\cite{kraus2019impact, arms1999benefits, nelson1999xgobi}, and outlier detection~\cite{wagner2018immersive} were performed more accurately in an IE than the traditional 2D desktop setting. However, immersive analytics has a trade-off between task execution time and learning curve with accuracy. The studies argue that `true' 3D representation could increase users' perception of the structure of 3D visualizations in immersive analytics. From a different perspective, Etemadpour et al.~\cite{etemadpour2013effect} and Arns et al.~\cite{arms1999benefits} report that immersive analytics could encourage users to interact with data with a greater sense of immersion, and it leads to higher accuracy in their study tasks. 

Furthermore, earlier studies present formal user studies performing visual analytics tasks in a 2D desktop and an IE to compare performances. Various visualization types including maps, networks, scatter plots, and parallel coordinates plots are evaluated and considered promising for immersive analytics. More details could be found in earlier survey works~\cite{kraus2022immersive, fonnet2019survey, frohler2022survey}. 

These earlier studies highly expand our understanding of using visualizations in IEs.
For example, Kraus et al.~\cite{kraus2019impact} point out that users could lose their spatial orientation when they are in huge immersive visualizations. They suggest providing a limited size of 3D data visualizations instead of surrounding users with data points. 
However, only a few studies have looked at 3D user interaction techniques to support users' active immersive analytics tasks in IEs. 
Data exploration for immersive analytics is performed by moving around a given 3D space or repositioning 3D plots, unlike traditional 2D desktop environments. This highlights the need to understand how well 3D user interaction methods could support human spatial ability. 
Our work evaluates 3D user interaction techniques for their spatial working memory support to explore data in VR and AR environments using 3D scatter plots. 

\subsection{Interaction Techniques for Immersive Analytics}\label{sec_relatedWork2}

Natural and intuitive user interactions are essential for users to do effective immersive analytics~\cite{slay2001interaction, belcher2003using, marsh2013cognitive}. 3D UI Researchers 
have introduced a number of interaction techniques that mimic real-world affordance or surpass its limitations in order to support the user's selection, manipulation, and navigation in an IE~\cite{al2018virtual}. Among the 3D UI techniques, earlier visualization studies evaluated Walking, World Grab~\cite{robinett1992implementation}, Word-In-Miniature (WIM)~\cite{stoakley1995virtual}, and Teleportation for immersive analytics navigation. The Walking technique is a first-person perspective-based locomotion technique for navigating in an IE. It includes physical movement, arm-swing, and walking-in-position that walk and rotate in a position~\cite{slater1995virtual}, and glider metaphor~\cite{mine1997moving, mine1995virtual} that uses controllers. World Grab shifts the location of visualization for 3D data exploration instead of moving users' locations. WIM presents a miniature version of a 3D plot to users. WIM allows users to interact with data objects and change the plot orientation by interacting with the miniature. Teleportation is the most commonly used navigation technique in VR. Users can instantly travel to a distant location by directing a controller's ray toward that location and pressing a button.

The previous studies evaluate these navigation techniques for performing visual analytic tasks in an IE~\cite{usoh1999walking, bowman1999maintaining}. The tasks include identifying the highest degree node and neighbors of the node, counting the number of specific nodes, and finding links that connect two nodes from 3D network visualizations. Henry et al.~\cite{henry2010effects} compared egocentric (e.g., flight metaphor) and allocentric (e.g., World Grab) navigation techniques for network visualizations in a Cave Automatic Virtual Environment (CAVE)~\cite{cruz1993surround}.
They found no difference between the two navigation techniques for performing the identification tasks. Zielasko et al.~\cite{zielasko2016evaluation} compared five different walking metaphors while seated.
The five walking metaphors include walking-in-position, arm-swing, foot pedal metaphor, forward and backward leaning against a chair, and gamepad control (gliding). They report that the body-leaning and the pedal metaphor performed the best for node-link identification tests while sitting. Drogemuller et al.~\cite{drogemuller2018evaluating} compare four navigation techniques (two flight methods, teleportation, and WIM) for graph analysis. They argued that WIM would be the most efficient solution to complete the tasks because it gives users a better understanding of an overview of a 3D graph.

Simpson et al.~\cite{simpson2017take} and B{\"u}schel et al.~\cite{buschel2017investigating} investigate the effect of 3D user interaction techniques on users' high-level visual analytics tasks such as clusters and trends in IEs. They compare Working and World Grab navigation on 3D scatter plots in VR and AR environments, respectively. 
They found no significant difference between the two techniques for spatial awareness of the 3D scatter plots. 
Wagner et al.~\cite{wagner2021effect} and Lages and Bowman~\cite{lages2018move} investigate the effect of Walking and Grab metaphors on search and comparison task performance in an immersive analytics setting. They reported in common that Walking helps users with better spatial abilities than Grab. 
However, it is still unclear how effectively 3D user interaction techniques support users' low-level tasks such as memorizing and recalling individual data for 3D data exploration. 
The ability to recall data based on spatial information, called `spatial working memory,' is as important as the ability to comprehend visualization structures. Our work assesses the effects of three 3D user interaction techniques on spatial working memory in the context of data exploration in VR and AR. 

\section{Methods}

This section introduces our study methods including a task, a 3D plot setting, and 3D user interaction techniques. Our study stems from a research question, \textit{\textbf{``Which 3D user interaction technique would be the most effective for users to explore 3D visualizations for immersive analytics?''}}
3D user interaction techniques support users' data exploration in immersive analytics. 
Understanding their spatial working memory support is important to do low-level visual analytics tasks~\cite{amar2005low, brehmer2013multi} like retrieving and comparing values and work on complex tasks while keeping the information in mind. 

\subsection{Study Design and Corsi block-tapping Test}
The Corsi block-tapping task is a psychological test for assessing visuo-spatial working memory. Visuo-spatial working memory indicates the capacity to maintain visuo-spatial information for a short term~\cite{rizzo2002psychoanatomical}. The original task displays 9 square blocks in a 2D display and shows a user a specific sequence of blocks by highlighting them one by one. Following that, the user is instructed to select the blocks in the same order. The number of block sequences increases from two to up to nine until the user fails to imitate the sequence. This number is known as the Corsi Span, and its average is 5 or 6.

Our study is designed based on the Corsi block-tapping task to evaluate 3D user interaction techniques on visuo-spatial working memory with the context of data exploration within an immersive 3D space. 
It displays thirty 3D sphere blocks that are considered as data points for a 3D scatter plot. Our task could be thought of as a low-level scatter plot operation to search for and locate an object~\cite{sarikaya2017scatterplots, casner1991task}.  
We limit the number of block sequences to three because our goal is to evaluate interaction techniques rather than an individual's visuo-spatial working memory ability. 
We decided to use fewer sequences than the average of the Corsi span as our task is more complex than the original task. 
In addition, the earlier research~\cite{vandierendonck2004working, van2014factorial} shows that brain activity for spatial analysis remains the same even when the number of block sequences to recall exceeds 3 or 4. 


\begin{figure}[t]
    \centering
    \includegraphics[width=.95\columnwidth]{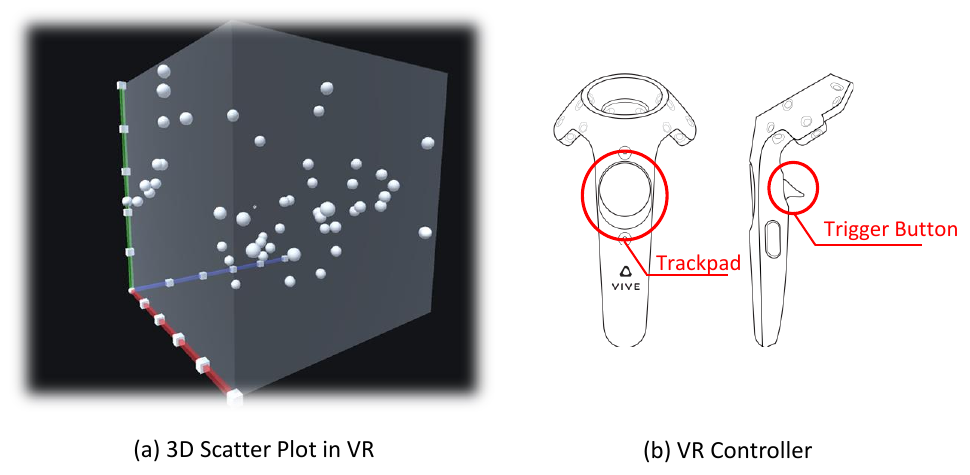}
    \caption{(a) 3D Scatter Plot in VR: we use the Iris dataset and each axis of the scatter plot represents one of four attributes of the dataset. (b) A Vive VR Controller is used for interaction.  
    }
    \label{fig:fig_vr_setting}
\end{figure}

\subsection{3D Scatter Plot}\label{subsection_scatterPlot}
A 3D scatter plot with 30 data points is used as a stimulus in our tapping task. We choose the 3D scatter for the task stimulus because it requires participants to search for and recall a point using spatial information (i.e., x, y, and z coordinates). 
While our study's focus is not on the specific type of dataset and a range of real-world and populated data can be applied for the study, we use Iris dataset\cite{IRIS} to draw the 3D scatter plot because it is popular and easily accessible in data exploration and visualization. The dataset contains 150 data records, each with 4 attributes (sepal length, sepal width, petal length, and petal width). For each task trial, the 30 data points and their 3 attributes are randomly selected from the 150 records and the four attributes respectively. In order to avoid any potential clusters formed by 30 data points acting as a visual cue in remembering the location of the data points, we randomly select three values out of four features for each trial. After the 30 records are selected, the 3 attribute values are normalized according to each axis. Fig.~\ref{fig:fig_vr_setting}-a shows an example of the 3D scatter plots.
The x-, y-, and z-axes are represented in red, green, and blue bars respectively. The origin (i.e., \textit{$<x,y,z>$) = $<0,0,0>$}) is the position where the three bars converge. The spheres are white by default, and they are not casting shadows on each other.

The 3D scatter plot for our study has limited sizes to avoid for users to lose their orientation in an IE, as Kraus et al.~\cite{kraus2019impact} suggested.
The size of the 3D scatter plot for the VR study (Fig.~\ref{fig:fig_vr_setting}-a) is 0.5m (width) \texttimes{} 0.5m (depth) \texttimes{} 0.5m (height).
The spheres in the scatter plot represent data points of the 3D scatter plot. They are used as the blocks for our tapping task. Their radius is 2.5 cm. For the AR study, however, we resize it due to the limited field of view of our AR device. Its size is 0.3m \texttimes{} 0.3m \texttimes{} 0.3m (Fig.~\ref{fig:comparison2}-a). The radius of data points is also reduced in the same ratio, having 1.5cm. 

\subsection{3D User Interaction Techniques}
We investigate three interaction techniques in this work. The techniques allow users to navigate the 3D plot along its side.

\textbf{Walking:} 
Users can physically walk within IEs to navigate a 3D plot.  
This is the most natural way of traveling because of the high biological symmetry of how humans move in the real world, but it requires a physical space in the real world that has to be the same size as the 3D space.

\textbf{Teleportation:} 
Teleportation is widely used in many VR applications. If the user points to a floor position by using an input device and activates Teleportation, the user will instantly move to that location~\cite{bowman1997travel}. It allows users to navigate an IE while standing in real-world space. In this work, there are two pre-defined areas for Teleportation for the target locations for 90\degree and 180\degree view transition from the current position. 


\textbf{Grab:} 
This is similar to World Grab in Section~\ref{sec_relatedWork2} which can control 6 Degrees of Freedom (DOF, x, y, z + yaw, pitch, roll). In this work, however, Grab only allows users to rotate 3D plots along its y-axis (yaw). Users can grab a 3D plot by operating one controller or hand and changing its orientation to see the plot from a different perspective. 
Unlike Walking, users are not allowed to walk around in both an IE and real-world space, but they can move their torso for maneuvering in the same location. 

In terms of the 3D UI perspective, Walking and Teleportation are the navigation (locomotion) techniques, while Grab is an object manipulation technique. However, the user may perceive Grab as a navigation technique like the scene-in-hand approach \cite{ware1990exploration}. 
In the following sections, we present two comparative studies in VR and AR. The VR study evaluates all three techniques. However, since Teleportation is not applicable in AR, we evaluate only Walking and Grab in the AR study. The specifics of how to operate these interactions in VR and AR are explained in the following sections, along with VR/AR device descriptions.


\section{Study 1: Spatial Working Memory in VR}

\begin{figure*}[t]
\centering
\includegraphics[width=.90\linewidth]{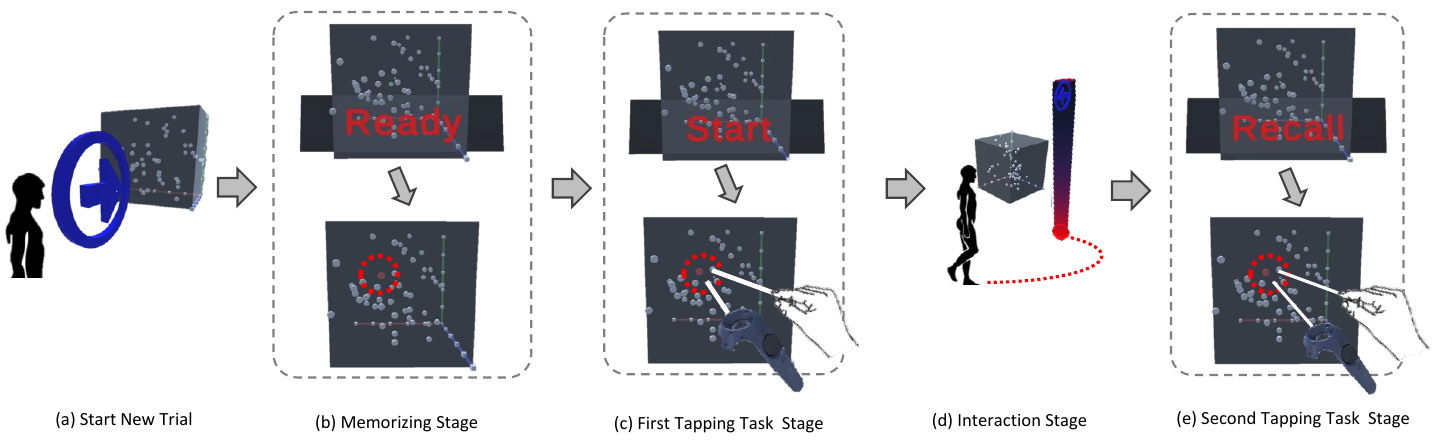}
\caption{Task Procedures. (a) As a trial begins, a participant moves to a designated position where an arrow marker is located. (b) In the memorizing stage, three data points are highlighted. The participant is asked to remember their positions and order. (c) The participant completes the first tapping task in the same position. (d) The participant is asked to navigate IEs or manipulate the 3D plot for data exploration. (e) At the new position, the participant completes the second tapping task recalling the data points and their order.
} 
\label{fig_task_procedure}   
\vspace{-0.5cm}
\end{figure*}

\subsection{Participants}
We recruit 18 participants (12 females and 6 males, 17 right-handed and 1 left-handed, average age: 21, ranging from 19 to 22.) from the student participants recruitment system (SONA) in our university. 
All participants have 20/20 (or corrected 20/20) vision and have no impairments in the use of VR devices. They rewarded 0.5 SONA credit as compensation for their participation.

\subsection{Procedures}
The study takes approximately 45 minutes (IRB \#13077). 
Upon arrival, a participant signs the informed consent form and completes a demographic questionnaire. Then the participant is briefed on the study objective and procedures, three interaction techniques (i.e., Walking, Teleportation, and Grab), and the task. In the training session, the participant practices the task twice per technique to become acquainted with the task and techniques.

The task consists of four stages (Fig.~\ref{fig_task_procedure}):  1) the memorizing stage, 2) the first block-tapping stage, 3) the interaction stage, and 4) the second block-tapping stage sequentially. 
In the beginning, the participant is asked to move to a start position where an arrow mark is located (Fig.~\ref{fig_task_procedure}-a). 
It is positioned 75 cm away from the center of the scatter plot. 
In the position, the participant can see entire scatter plot.
When the participant approaches within 0.1 m of the center of the arrow mark, the arrow mark turns red, indicating that he/she is ready to begin a trial. 
Please note that the 3D scatter plot and arrow mark are all set at the height of the participant's eye height minus half the plot height. 
When the participant clicks the trigger button, the memorizing stage is started. 

In the memorizing stage, the participant is asked to memorize the positions and order of three data points being highlighted. As shown in Fig.~\ref{fig_task_procedure}-b, a panel \textit{Ready} is popped up and then disappears in a second. Following that, three out of 30 data points are highlighted in red for 2 seconds one after the other. Three data points are randomly selected while considering not to be occluded to other data points from the participant's view. Next is the first block-tapping stage. 

The first block tapping stage starts with a panel \textit{Start} (Fig.~\ref{fig_task_procedure}-c) appearing for 1 second after the memorizing stage. When the panel disappears, the participant is asked to select the three data points in the sequence order as precisely as possible. 

Once the participant selected the three data points, next is the interaction stage (Fig.~\ref{fig_task_procedure}-d). 
The participant navigates the plot to a target position using one of three interaction methods. 
The target position is marked as a column with the arrow as shown in Fig.~\ref{fig_task_procedure}-d. 
Its position is constant at 0.75m from the center of the plot. 
If the user's dominant hand is the right hand, the target location is to the left (90\textdegree{} movement) or opposite side (180\textdegree{} movement) of the plot to the current user's position. Conversely, if the dominant hand is left, the target position to move is the right (90\textdegree) or opposite (180\textdegree) side of the plot to the user’s position. We set the left and right-hand settings differently for the 90\textdegree{} movement case because their wrist rotation directions differ for the Grab interaction. The arrow mark at the target position functions in the same way that it does at the start position.

At the target position, the participant can move on to the second block-tapping stage by clicking the trigger button. In the second block-tapping stage (Fig.~\ref{fig_task_procedure}-e), the participant recalls the three data points in the order which he or she saw in the memorizing stage. These four stages are one task trial in our study, and it allows us to evaluate the three techniques' spatial-memory support and interaction time to explore 3D data in IEs.

The participant performs the task 30 times (3 interaction techniques \texttimes{} 2 angles of view transition \texttimes{} 5 trials). The participant is asked to perform the task as accurately as possible. Sequences of the interaction techniques are counterbalanced across all the participants using a Latin square design. The view transitions angle is randomly ordered. After 10 trials with one technique, the participant takes off a VR headset and directs to a desktop to evaluate a navigation method subjectively by answering questionnaires.
After all the tasks, the participant is lastly asked to complete a post-questionnaire. The post-questionnaire asks to rank the three techniques from 1 (best) to 3 (worst) according to their preference for interaction methods.

\subsection{Apparatus and Controller Operations}
We used a Vive Pro 2 Head Mounted Display (HMD) with a Wireless Adapter, one Vive controller, and Vive Base Station 2.0. 
The headset has a 120\textdegree{} Field-of-View, a resolution of 2448 \texttimes 2448 for each eye, and a 90 Hz refresh rate in the wireless setting. The VR application for the study is implemented in Unity 2020.3.32f1 and ran on a Windows 11 desktop which has Intel Xeon W-2245 CPU (3.90GHz), 64GB RAM, and Nvidia GeForce RTX 3090 graphics card. 

Teleportation and Grab require controller operations while Walking does not. We use a Vive controller (Fig.~\ref{fig:fig_vr_setting}) as an input device. For Teleportation, when the user presses the trackpad on the controller, a ray emerges from the controller in VR. 
The ray falls to the floor in an arc shape, showing a ring at their intersection (Fig.~\ref{fig:teaser}-d). 
When the user points the ray at the area and releases the thumb from the trackpad, it transfers the user to the target location with short fade-in and out effects. 
After Teleportation, the orientation of the user is automatically towards the 3D scatter plot.

For Grab, the user can grab and rotate a 3D scatter plot. Grabbing the plot can be achieved by placing the controller nearby the 3D scatter plot and clicking the controller’s trackpad. The user can then rotate the plot along the y-axis (yaw) by turning the controller with his or her wrist. 
The target position (i.e., the column with the arrow in Fig.~\ref{fig_task_procedure}-d) also rotates following the plot’s orientation.

During the tapping stages, a ray is emitted from the controller.
The participant can select a sphere by clicking its trigger button.  
The trigger button is also used to progress to the next stages of a task. 

\subsection{Measurements and Hypothesis}\label{subsection_measurements}
For the study, we collect the following measurements to evaluate the 3D user interaction techniques for spatial working memory.

\textbf{Distance Error:} It assesses how well a participant recalls data points in the correct order during the two block-tapping stages. It is measured by taking the Euclidean distance between the center of a user-selected data point and the center of its ground truth data point. The distance error is measured for each tapping order in the three data sequences.
Please note that the distance error could include the two radii of data spheres.
A lower error indicates higher accuracy. 

\textbf{Interaction Time:} It is the time taken for the interaction stage (Fig.~\ref{fig_task_procedure}-d) to explore a 3D scatter plot.


\textbf{Subjective Performance:} After the participant completes the task using one of three interaction techniques, subjective performances including physical demand for the interaction technique and degree of task success are evaluated. 
The two questions are rated on a 7-point Likert scale from 1 (Not At All) to 7 (Very Much).

\textbf{Motion Sickness:} VR motion sickness is a well-known problem that VR users encounter during their VR experiences. The degree of sickness is measured using the VRSQ questionnaire~\cite{kim2018virtual}. The questionnaire has 9 items including general discomfort, headache, and eye strain, and each item is evaluated on a 4-point Likert scale (0 (None) to 3 (Severe)). To measure the VR sickness score (range: 0$\sim$100), the following formula is used where \textbf{\textit{i}} refers to the question number in the VRSQ questionnaire, and \textbf{\textit{s(i)}} indicates each Likert scale score.
\vspace{-0.3cm}
\begin{equation} \label{VRSQ_eq}
VRSQ Score=((\sum_{i=1}^{4} s(i)) /12 *100 + (\sum_{i=5}^{9} s(i)) /15 *100 )/2
\vspace{-0.3cm}
\end{equation}

Our main hypotheses are as follows. 
\begin{enumerate}[label=H\arabic*. , noitemsep, leftmargin=*]
\item Walking and Grab would have greater accuracy than Teleportation. This is due to the fact that both Walking and Grab allow the user to track how data points in the 3D scatter plot are transitioned, while Teleportation does not.

\item Teleportation would have the shortest interaction time because it allows the user to move instantly to a target location.


\item Teleportation would cause more severe motion sickness than Walking and Grab because it causes an abrupt view change and disorientation~\cite{bowman1997travel, bowman1999maintaining}.

\item On a second tapping task trial, distance errors will increase with tapping order because the initially presented items are most effectively stored in memory, called the primacy effect~\cite{deese1957serial, murdock1962serial}.
\end{enumerate}







\subsection{Results}

Multiple statistical analyses, including between-subjects and within-subjects Analysis of Variance (ANOVA) tests at the 95\% confidence level, are used. 
The measurement of distance errors at the first tapping task is analyzed using between-subject ANOVA regarding the participants. Distance errors at the second tapping task are analyzed using the three-way repeated measures ANOVA test regarding interaction techniques, view transition angles, and tapping order. 
Next, two-way repeated measures ANOVA is used to examine interaction time in terms of the interaction techniques and view transition angles. We find that interaction time violates normalcy but nevertheless meets sphericity. We opt to use ANOVA because a previous study~\cite{blanca2017non} provided empirical evidence for its robustness when evaluating data with non-normal distributions. 
The post-hoc analysis is conducted with a Tukey's HSD at the same 95\% confidence level with a Bonferroni adjustment.

The subjective performance variables and motion sickness score are analyzed with the one-way ANOVA test and the Tukey's HSD post-hoc test at the familywise 95\% confidence level regarding the navigation techniques. Please note that the ANOVA test can be conducted on Likert-type data if a Likert item contains at least 5 categories~\cite{harpe2015analyze, hsu1969effect, norman2010likert}. In the case of the subjective performance variables, they are measured using the 7-Likert scale and so ANOVA is applicable. The VRSQ score is a continuous variable calculated as the sum of nine four-point Likert-scale items. As a result, it could also be analyzed using ANOVA.

    

\begin{table}[ht]
\small
\caption{\label{table_result_s1_second_tapping} \textbf{Results on Distance Error at Second Tapping Task in VR}}
\begin{tabular} {m{4.0cm} | m{0.8cm} | m{0.5cm}| m{0.6cm} | m{0.5cm}}

\toprule
    {\textbf{Factor}} & {\textbf{$DOF$}}  & {\textbf{$F$}} & {\textbf{$p$}} & {\textbf{$\eta$\textsubscript{$p$}\textsuperscript{$2$}}}\\ \midrule
    Technique \texttimes{} View Transition Angle  & {2, 178} & {.171}  & {.843} & {.002} \\ 
    Technique \texttimes{} Tapping Order & {4, 356} & {1.09}  & {.358} & {.012} \\ 
    View Transition Angle \texttimes{} Tapping Order & {2, 178} & {1.53}  & {.218} & {.017} \\ 
    
    Technique  & {2, 178} &  {5.17}  & {.007} & {.055} \\
    View Transition Angle & {1, 89}  & {6.68} & {.011} & {.070} \\
    Tapping Order & {2, 178}  & {7.29} & {\textless.001} & {.076}\\
    \bottomrule
\end{tabular}
\vspace{-.3cm}
\end{table}

\textbf{Distance Error}
We first report the results of the first tapping task.
We find no significant difference between the participants on distance error in the first tapping task before interaction ($p$=.548, $F(17, 522)=.922$, ${\eta_p}^2=.029$, $m$=.042 m, $SD=.118$). 
This result implies that the participants have comparable spatial abilities at least at the first tapping task before interaction. Please note that this does not mean that all participants have the same spatial ability.

\begin{figure}[t]
\centering
\includegraphics[width=.95\columnwidth]{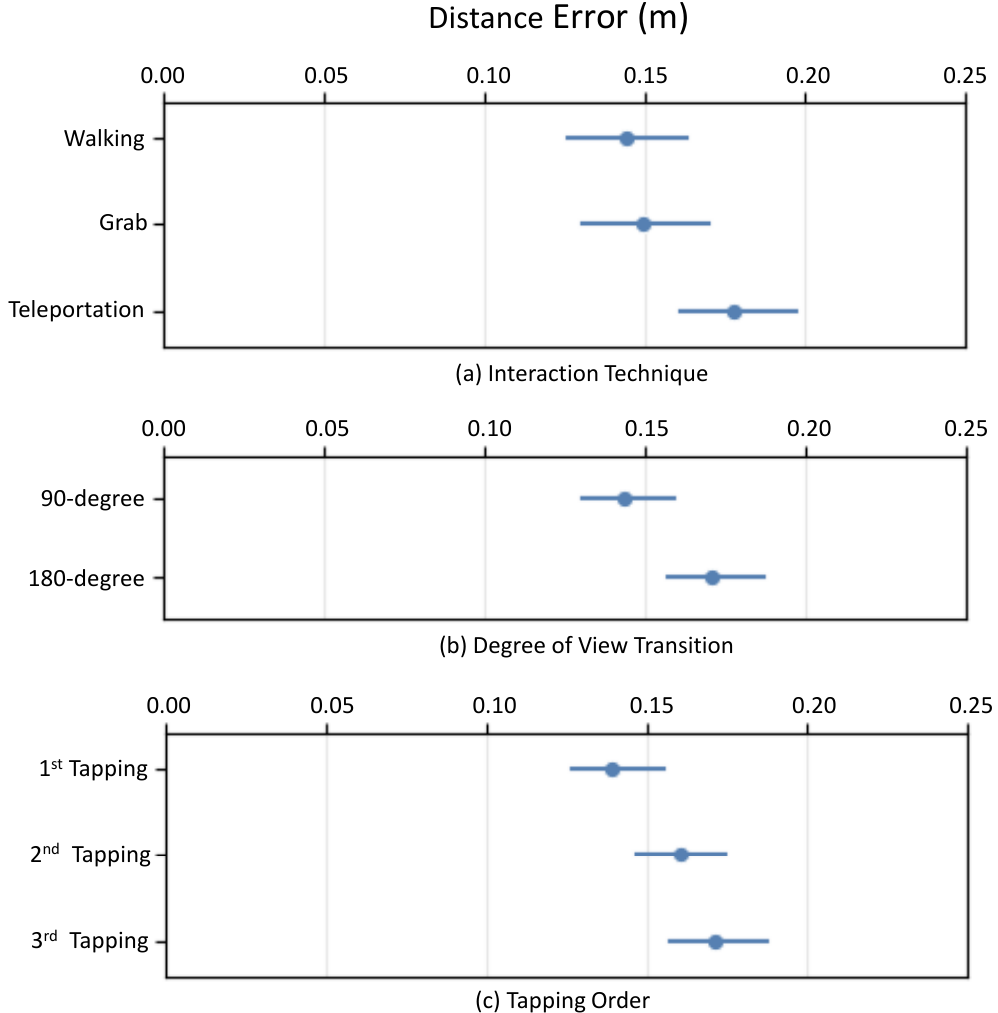}
\caption{Distance Error Results with 95\% CI in VR.
} 
\label{fig_result_vr_error_distance} 
\end{figure}

The statistical results in the second tapping task are reported in Table~\ref{table_result_s1_second_tapping}. 
There is a main effect on the techniques ($p$=.007). The pairwise comparisons show that Teleportation ($m$=.178 m, SD=.127) has a larger distance error than Walking ($m$=.143 m, SD=.130, $p$=.009) and Grab ($m$=.149 m, SD=.135, $p$=.041, Fig.~\ref{fig_result_vr_error_distance}-a).  However, there is no difference between Walking and Grab ($p$=1.00). This supports H1. There is also a main effect on view transition angles ($p$=.011). 
 90\degree{} transition angle ($m$=.143 m, SD=.132) has a significantly smaller distance error than 180\degree{} transition ($m$=.170 m, SD=.131, $p$=.011, Fig.~\ref{fig_result_vr_error_distance}-b). We also find a main effect on the tapping orders ($p$\textless.001). The first tap ($m$=.139 m, SD=.166) has a statistically smaller distance error than the second ($m$=.160 m, SD=.169, $p$=.024) and third ($m$=.171 m, SD=.178, $p$=.005) tapping shown in Fig.~\ref{fig_result_vr_error_distance}-c. No statistical difference is found in the second and third tapping ($p$=.481). These findings partially support H4.

\begin{figure}[t]
\centering
\includegraphics[width=.93\columnwidth]{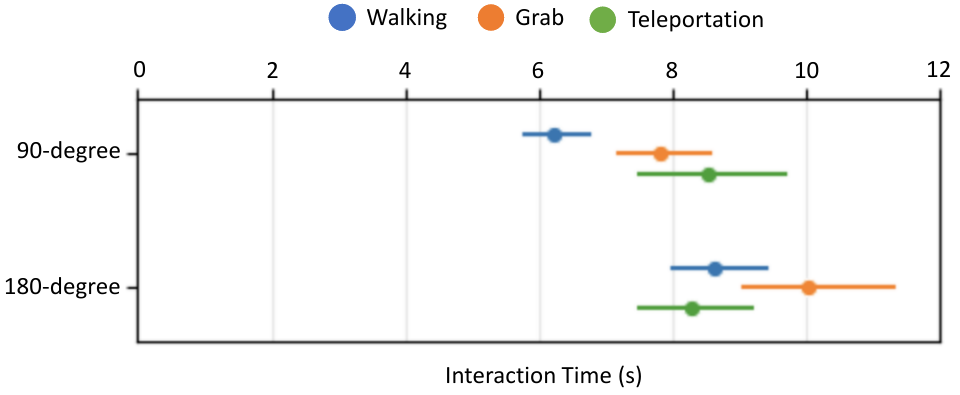}
\caption{Interaction Time Results with 95\% CI in VR. 
} 
\label{fig_result_vr_time}  
\vspace{-.4cm}
\end{figure}

\textbf{Interaction Time} 
The results have a significant interaction effect between interaction techniques and view transition angles (F(2,178), $p$\textless.001, $\eta$\textsubscript{$p$}\textsuperscript{$2$}=.100). We find simple effects on the 3D user interaction technique in 90\degree{} ($p$\textless.001) and 180\degree{} ($p$\textless.001) view transition conditions(Fig.~\ref{fig_result_vr_time}). 
In the 90\degree{} transition condition, Walking ($m$=6.23 s, SD=2.27) has faster interaction time, than Grab ($m$=7.8 s, SD=3.5,$p$\textless.001) and Teleportation ($m$=8.51 s, SD=5.48, $p$\textless.001). However, there is no difference between Grab and Teleportation ($p$=.750). In the 180\degree{} condition, Teleportation ($m$=8.26 s, SD=4.16) has statistically faster interaction time than Grab ($m$=10 s, SD=5.27, $p$=.006), and Walking ($m$=8.6 s, SD=3.35) is also significantly faster than Grab ($p$=.032). There is no difference between Walking and Teleportation ($p$=1.00). Simple effects of the view transition conditions in Walking ($p$\textless.001) and Grab ($p$\textless.001) are found. For the Walking and Grab interactions, the 90\degree{} transition has a faster interaction time than the 180\degree{}. We find no simple effect on Teleportation. This partially supports H2.

\textbf{Subjective Analysis} 
Significant differences between interaction techniques are found ($p$=.008, $F(2, 34)=5.64$, ${\eta_p}^2=.249$). Teleportation ($m$=1.44, SD=1.04) has statistically a lower physical requirement than Walking ($m$=2.22, SD=1.16, $p$=.035). However, no significant difference is found between Teleportation and Grab ($m$=1.89, SD=1.18, $p$=.311).

Next, the self-success results are presented. The significant differences between the three interaction techniques are disclosed ($p$\textless.001, $F(2, 34)=9.56$, ${\eta_p}^2=.360$). The degree of self-success with Walking ($m$=4.61, SD=1.50) is higher than with Teleportation ($m$=2.94, SD=1.34, $p$=.003), but no difference between Walking and Grab ($m$=4.22, SD=1.39, $p$=.823). Grab and Teleportation have significant differences as well ($p$=.024). 

\textbf{Motion Sickness} No difference between the three interaction techniques is reported ($p$=.878), rejecting H3. 
The average sickness scores for Walking, Teleportation, and Grab are 51.5 (SD=17.1), 52.2 (SD=18.6), and 52.8 (SD=19.5) out of 100 respectively. 


\section{Study 2: Spatial Working Memory in AR}

Study 2 evaluates the usability and effectiveness of two interaction techniques (i.e., Walking and Grab) in AR. 



\subsection{Participants}
A total of 10 participants (1 female and 9 males, average age = 22.9, ranging from 20 to 26) were recruited from SONA. They are all right-handed and have 20/20 (or corrected 20/20) vision. 9 participants reported that they have previously used AR applications like Pokemon Go~\cite{PokemonGo}.
The participants received 0.5 SONA credits for their participation. No participant participated in Study 1.

\subsection{Procedures}
Study 2 is a 30-minute-long study (IRB \#13239). 
The overall procedures and tasks are similar to Study 1. The differences are the size of the 3D scatter plot as we mentioned in Section~\ref{subsection_scatterPlot} due to the Hololens 2's limited field-of-view, and the limited number of supporting interaction techniques. We exclude Teleportation because it could not be achieved in an AR environment as a navigation method. Thus, Study 2 evaluates only Walking and Grab. 

In the study, a participant performs a total of 20 tasks (2 techniques \texttimes{} 2 view transition angles \texttimes{} 5 trials). The participant is asked to complete the task fast and accurately. After completing the task using one of the interaction techniques, the participant performs a subjective evaluation of one technique and the motion sickness questionnaire. After finishing all the tasks, the participant completes a post-questionnaire.

\subsection{Apparatus and Controller Operations}

HoloLens 2 is utilized for Study 2. It is an AR HMD having a 50\textdegree{} Field-of-View, a resolution of 2048 $\times$ 1080 for each eye with a 75Hz refresh rate. The task application is built and executed on the same Unity version and desktop used in Study 1. When the application is played on the desktop, holographic content is streamed to HoloLens 2 in real-time through Holographic Remoting Player. Holographic Remoting Player is a HoloLens 2 built-in application that supports real-time streaming of desktop content to HoloLens 2.

\begin{figure}[t]
    \centering
    \includegraphics[width=.9\columnwidth]{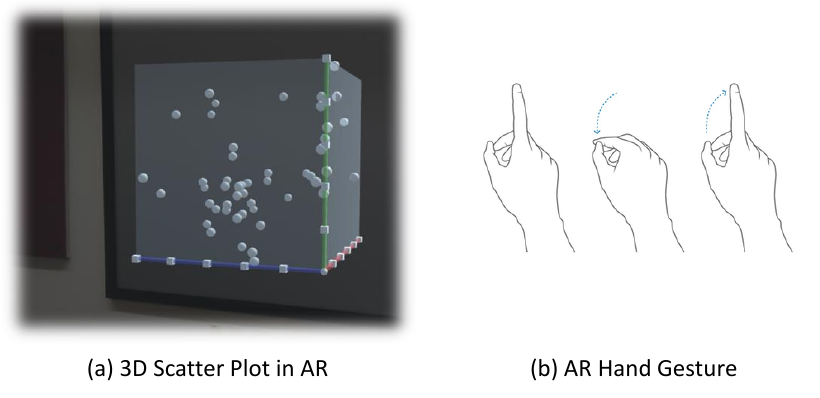}
    \caption{(a) 3D Scatter Plot in AR: users can see real-world backgrounds beyond the plot. (b) The AR hand gesture is also known as Air Tap.
    A ray is emitted from a user's hand, and the user can use the pinch gesture to interact with a virtual object that the ray has hit. }
    \label{fig:comparison2}
    \vspace{-.2cm}
\end{figure}

Unlike the VR device in Study 1, HoloLens 2 does not have a controller.  It supports a hand-tracking technique, and users can interact with AR objects by hand gestures. In our study, the participants can interact with 3D scatter plots with a pinch gesture (Fig.~\ref{fig:comparison2}-b). 
For Grab, users can pick up a 3D plot with a single pinch and make it rotate by turning their wrist. The single pinch is also used to select data points from a 3D scatter plot in the tapping stages. A double pinch is utilized to move on to the next stage of a task.

\subsection{Measurements and Hypothesis}
 Same as Study 1, we measure distance error, interaction time, tapping task completion time, subjective performance, and motion sickness score (See Section ~\ref{subsection_measurements}). Though the VRSQ questionnaire is designed for measuring the degree of motion sickness in VR, we use it to evaluate the motion sickness of the interaction techniques in AR. The hypotheses for Study 2 are as follows.

\begin{enumerate}
[label=H\arabic*., noitemsep ,leftmargin=* ]
\setcounter{enumi}{4}
\item Walking and Grab would have similar distance errors as both allow users to track the 3D scatter plot is transitioned.

\item Walking would have less interaction time than Grab as it does not require additional gestures to explore data in AR.


\item Walking and Grab would have a similar degree of motion sickness because both allow users to see smooth transitions of the scatter plot by following users' actions.

\item On a second tapping task trial, a later ordered tapping  would have a greater distance error than an earlier order as the same reason for H4.

\end{enumerate}

\subsection{Results}
The same statistical analytic methods as Study 1 are used. 

\begin{table}[ht]
\small
\caption{\label{table_result_s2_second_tapping} \textbf{Results on Distance Error at Second Tapping Task in AR}}
\begin{tabular} {m{4.0cm} | m{0.8cm} | m{0.5cm}| m{0.6cm} | m{0.5cm}}

\toprule
    {\textbf{Factor}} & {\textbf{$DOF$}}  & {\textbf{$F$}} & {\textbf{$p$}} & {\textbf{$\eta$\textsubscript{$p$}\textsuperscript{$2$}}}\\ \midrule
    Technique \texttimes{} View Transition Angle  & {1, 49} & {.406}  & {.527} & {.008} \\ 
    Technique \texttimes{} Tapping Order & {2, 98} & {.920}  & {.402} & {.018} \\ 
    View Transition Angle \texttimes{} Tapping Order & {2, 98} & {.331}  & {.719} & {.007} \\ 
    
    Technique  & {1, 49} &  {6.09}  & {.017} & {.111} \\
    View Transition Angle & {1, 49}  & {5.11} & {.028} & {.094} \\
    Tapping Order & {2, 98}  & {1.37} & {.257} & {.027}\\
    \bottomrule
\end{tabular}
\end{table}

\textbf{Distance Error}
In the first tapping stage, no significant difference between the participants on distance error ($p$=.964, $F(9, 190)=.331$, ${\eta_p}^2=.015$, $m$=.024 m, SD=.071) is found, which implies there is no difference on the participants' spatial ability. 

\begin{figure}[t]
\centering
\includegraphics[width=.95\columnwidth]{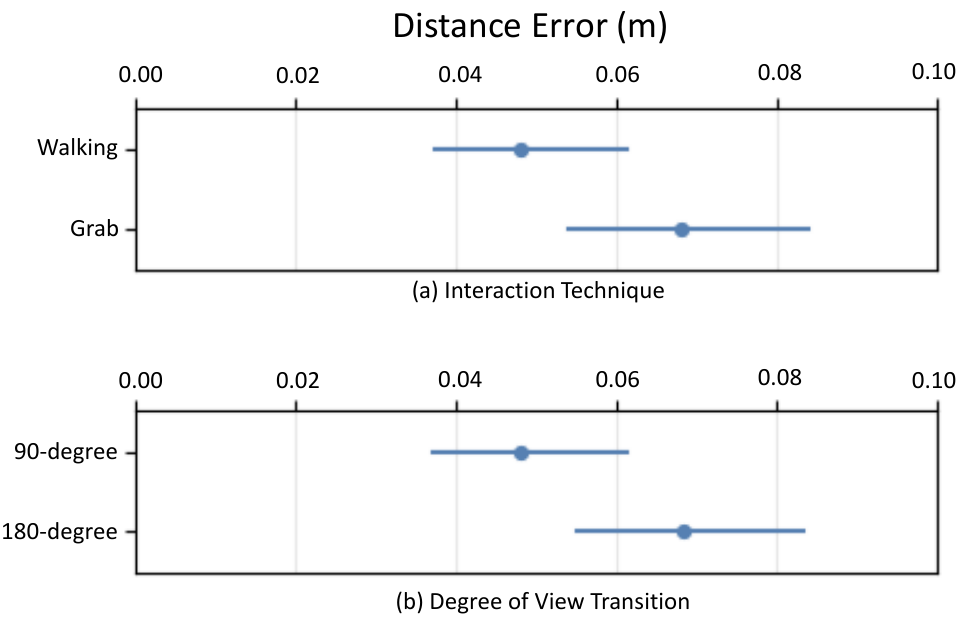}
\caption{Distance Error Results with 95\% CI in AR. 
} 
\label{fig_result_ar_error_distance}   
\vspace{-.2cm}
\end{figure}

Table~\ref{table_result_s2_second_tapping} presents the statistical findings for the second tapping task. 
No interaction effect is reported. 
There is a main effect of the techniques ($p$=.017, Fig.~\ref{fig_result_ar_error_distance}-a). Walking ($m$=.048 m, SD=.079) has a smaller error than Grab ($m$=.068 m, SD=.097). This rejects H5. There is also a main effect of view transition angles ($p$=.028, Fig.~\ref{fig_result_ar_error_distance}-b). 90\degree{} transition  ($m$=.068 m, SD=.097) has smaller distance error than 180\degree{} ($m$=.048 m, SD=.079). 
No main effect of the tapping orders is found ($p$=.257), rejecting H8. 

\begin{figure}[t]
\centering
\includegraphics[width=.95\columnwidth]{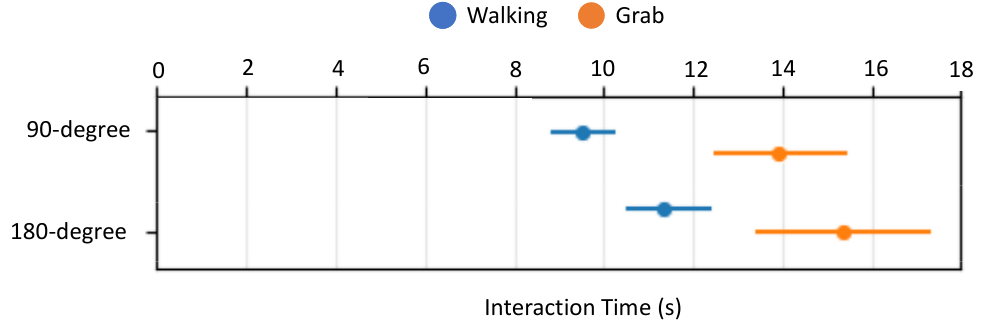}
\caption{Interaction Time Results with 95\% CI in AR.
} 
\label{fig_result_ar_time}   
\vspace{-.4cm}
\end{figure}

\textbf{Interaction Time}
We found no significant interaction effect between the navigation techniques and view transition angles ($p$=.751). Main effects of the techniques ($p$\textless .001) and view transition angles ($p$=.007) are found. 
Walking ($m$=10.4 s, SD=3.18) takes faster time to navigate than Grab ($m$=14.6 s, SD=6.05). This finding supports H6. 90\degree{} view transition ($m$=11.7 s, SD=4.69) has a significantly faster interaction time than 180\degree{} ($m$=13.3 s, SD=5.68). See Fig.~\ref{fig_result_ar_time}.

\textbf{Subjective Analysis}
Statistical results show that Walking ($m$=2.88, SD=1.28) and Grab ($m$=3.11, SD=1.52) have no significant effects on physical requirements ($p$=.724). Similarly, Walking ($m$=4.67, SD=1.05) and Grab ($m$=4.00, SD=1.24) have no significant difference in the degree of self-success ($p$=.125).

\textbf{Motion Sickness }
No significant difference is disclosed ($p$=.239). 
 Walking and Grab scored 51.9 (SD=17.5) and 54.8 (SD=17.1) respectively. It supports H7. 
\section{Comparison between VR and AR}

In the earlier sections, we reported the results of the VR and AR studies. This section compares these outcomes. In order to avoid the multiple comparison problem, a non-null hypothesis statistical test (non-NHST) focusing on the sample mean and bootstrapped 95\% confidence interval (CI) approach~\cite{calmettes2012making, dragicevic2016fair} is used. To compare the VR and AR study results, we report the differences between the mean values of the study conditions first, followed by the associated 95\% CI in square brackets. A narrow CI indicates strong evidence to assess the differences, but a CI including zero implies greater  uncertainty regarding the sign of an effect.

The distance errors and interaction time are analyzed. For a tapping task, its index of difficulty (ID) is measured by calculating movement distance from an object to the target object divided by target object size according to Fitt's law~\cite{fitts1954information}. As stated in Section~\ref{subsection_scatterPlot}, both the physical size of a target and the average travel distance to tap the target shrank by the same ratio of 0.6 in the AR study compared to theirs in the VR study. This means that the IDs for our VR and AR research are the same. This rationale allows us to compare the results of the two studies. 
To compare the distance error results, we perform data pre-processing. 
We normalize the VR distance error by multiplying 0.6. 
In the case of interaction times, pre-processing is not required. This is because the tasks in both environments ask a participant to move in the IE or rotate the plots so that they can view a 90\degree{} or 180\degree{} rotated view of the current view, 75 cm away from the center of the plot.
We set the following hypotheses:

\begin{enumerate}
[label=H\arabic*., noitemsep, ]
\setcounter{enumi}{7}
\item The distance errors in VR will be much smaller than in AR. We expect this because VR users could focus more on 3D plots than AR users due to their background characteristics. Real-world objects around the plots might distract AR users.

\item Walking in VR and AR will not have a difference in interaction time. This is due to the fact that Walking is dependent on users' own ability and is not characterized by technical differences.

\item Grab in VR and AR will have significantly different interaction times due to their technical differences (i.e., controller vs. pinch gesture) in interacting with the plots. 

\end{enumerate}

\begin{figure}[t]
\centering
\includegraphics[width=.9\columnwidth]{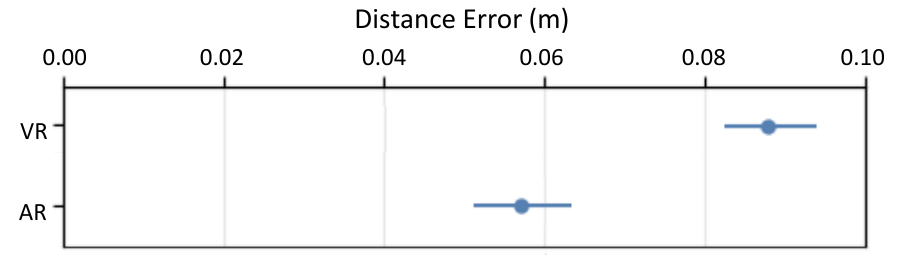}
\caption{Distance Error Results with 95\% CI in VR and AR. The VR results are normalized by multiplying the errors by 0.6.
} 
\label{fig_result_vr_ar_error}   
\end{figure}

\textbf{Distance Error} Fig.~\ref{fig_result_vr_ar_error} shows the bootstrapped sample means and their 95 CI in the VR and AR studies. The VR results show a larger error by .032m [.021, .041] compared to the AR results. This finding is evidence to reject our H8.

\begin{figure}[t]
\centering
\includegraphics[width=.95\columnwidth]{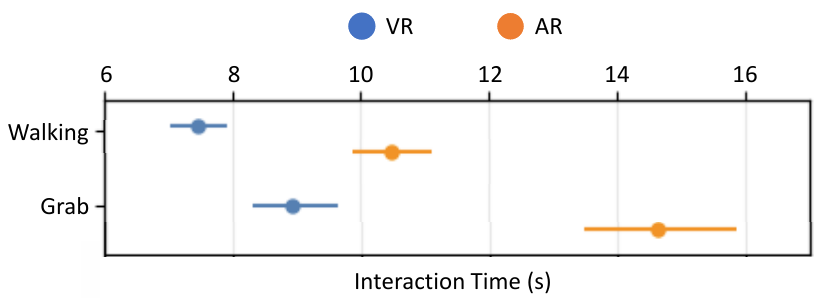}
\caption{Interaction Time Results with 95\% CI for Walking and Grab in VR and AR. 
} 
\label{fig_result_vr_ar_time}   
\vspace{-.4cm}
\end{figure}

\textbf{Interaction Time}
The VR and AR study results by the interaction techniques are shown in Fig.~\ref{fig_result_vr_ar_time}. It shows that Walking in VR is faster than Walking in AR by 3.03s [2.25, 3.81]. Similarly, Grab in VR is faster than Grab in AR by 5.72s [4.36, 7.08]. These findings support H10, but reject H9.
\section{Discussion}

In this section, we discuss our findings and limitations.  

\textbf{Walking outperformed other user interaction techniques in both VR and AR: }
Our results show that Walking in VR supports participants' mental rotation of the target points more accurately than Teleportation, having significantly smaller distance errors. 
In AR, we discovered that Walking accurately supports the participants in completing tasks more than Grab, whereas there is no difference between Walking and Grab in VR. 
These findings are consistent with Simpson et al.~\cite{simpson2017take}, but not with B{\"u}schel et al.~\cite{buschel2017investigating}, which compare Walking and Grab in spatial awareness with 3D scatter plots to perform high level analytic tasks (e.g., trend and cluster) in VR and AR settings respectively. 
Further research is needed to evaluate the effect of the interaction techniques on spatial ability on both high- and low-level visual analysis tasks in a systematic manner. 
In terms of interaction time, Walking is also significantly faster than other techniques in VR. Walking is only comparable to Teleportation for IE navigation when a larger movement is required. 


The participants' comments explain why Walking outperformed the others.
They stated that they prefer Walking as it could continuously monitor the transitions of data points in accordance with the viewpoint change while also allowing them to control the transition speed. 
Compared to Walking and Grab, they reported that Grab was difficult because they had to pay attention to their hands to rotate the plot, and made it hard to focus on two things (i.e., rotating the plot and remembering the sequence of data points) at once.

Interestingly, despite the fact that the results indicate that Walking helps participants complete the spatial memory task more accurately and quickly than Grab, participants' subjective performances on the two interaction techniques show no difference. The participants who stand on the opposite side of Walking, preferred Grab because it allows them to stand still and stay focused on the 3D scatter plot while they change the orientation of the plot. They said that Walking was inconvenient because it took their focus away from the data points while moving to the target position. 


\textbf{The participants in AR show better spatial memory performance than the participants in VR:} We expected that the real-world objects in AR would distract the participants and increase their cognitive load to memorize the correct data sequences. Interestingly, however, the result is very opposite to our hypothesis. In the VR study, the 3D plot is located in the center of the black room, and no virtual object exists around the plot. 
In contrast, the participants in the AR study can see their real hands and real-world objects that exist behind the virtual plot. 
Whitlock et al.~\cite{whitlock2020graphical} reported that users use their bodies or objects around virtual objects as referents to help with size, height, and depth estimation in AR. However, VR is lack such an embodiment. 
Contrary to our hypothesis, we now consider that environmental characteristics might have helped the participants to perceive the data positions as spatial anchors. Future research is required to understand the effects of surrounding real-world objects in AR. This would be applied to virtual objects in VR as well. 


Although the AR participants have a better accuracy than the VR participants, the VR participants completed the interaction faster than the AR participants. There could be two possible explanations. Firstly, we consider that environmental characteristics also have another effect on interaction time. 
We speculate that the real-world background in AR may have affected the participants to explore the 3D plots with caution. Secondly, particularly for Grab, we consider the different interaction ways (i.e., physical controller vs. hand gesture) for the VR and AR devices that may have an impact on the performances. The physical VR controller and physical walking provide solid haptic feedback and the sense of confirmation for their actions, but hand gesture does not. In addition, the VR device provides better input device tracking accuracy than AR~\cite{hubner2020evaluation, soares2021accuracy}. Three participants reported that the AR device would not recognize their inputs sometimes, and it made the task harder to grab the 3D scatter plot. One of them said \textit{``Grab just brought a lot of frustration as it did not seem to do what I wanted it to do.''} We expect this is due to their familiarity with the AR device. This issue is likely to be remedied when users become familiar with gesture operations or if a physical controller is adopted as an AR controller~\cite{buschel2019investigating, henrysson2007experiments}.


\textbf{Larger transition movements impair spatial memory performance: }
We find that the distance errors are increased along with the increased transition movements in both VR and AR. This finding can be explained in two ways. One potential reason is that the cognitive load on mental rotation increases with the size of the view transition required~\cite{shepard1986mental, shepard1971mental}. 
Secondly, occlusion might affect the participants' performances. As mentioned earlier, data points that are not occluded by other points from a participant's point of view are only allowed to be highlighted in the memorizing stage (Fig.~\ref{fig_task_procedure}-d). 
Those highlighted data points could be occluded by other points after the interaction stage. Differently speaking, the greater the required movement, the greater the possibility of occlusion.

\textbf{3D user interaction techniques have comparable motion sickness:} We hypothesized that Teleportation's sudden viewpoint change causes higher motion sickness than Walking and Grab. However, we found no difference among them, which is consistent with Rahimi et al.'s result ~\cite{rahimi2018scene}. 
We observed that many participants double-checked target data sequences before teleporting to target locations. We consider that these participants’ intense focus on the plots during the memorizing stage may have caused them to ignore the motion sickness. 

The participants, interestingly, reported that Teleportation requires less effort and fatigue to navigate an IE than Walking, and Grab though Teleportation does not support spatial-working memory well due to its instant view changes. 
Ten participants in VR Study commented that it is the simplest method for navigating in an IE. 
However, no one preferred Teleportation to Walking and Grab for the tapping task for the following reasons: 
In addition to its quick vision change, the participants said that activating Teleportation by looking at the destination resulted in the loss of visual tracking of the points. 
They additionally commented that they suffered from disorientation after teleporting to the target position and tried to figure out which angle of the points they were looking at.

\textbf{Highlighting/annotation techniques would help users in low-level data exploration in IEs.}
Our study results found that the error distance increased with the tapping orders in the VR study, though we found no difference in the AR study. Furthermore, it shows that the average distance errors in both studies increase before and after the interaction stage, although we did not report statistical results of them in this work. Given that the Corsi span in a 2D environment is 5 or 6, this study demonstrates that data exploration in IEs is a more difficult task. Based on these findings, we consider that providing highlight/annotation techniques to users would help them to explore 3D scatter plots more accurately.

\subsection{Limitation}
In our study, the effect of interaction techniques on spatial memory is investigated in a situation of navigating the hand-size 3D plot along its side views except for the top and down views. 
In practical terms, however, users might want to fully explore the 3D plots with 6 DoF manipulation.
Furthermore, if the size of a plot is large, he or she might even want to control the plot scale~\cite{yang2020embodied}. 
It requires up to an additional 3 DOF (x-, y-, z-axis scales) for the scale interactions. 
Previous research shows that techniques having more DOF result in lower performance~\cite{shovman2015twist, wickens1994implications, elmqvist2008rolling}. 
However, it is still unclear how these additional DOFs affect cognitive loads for users interacting with 3D plots and performing analytical tasks in IEs.

For the practical use of 3D user interaction techniques and visualizations~\cite{ens2021grand}, future research with more diverse interaction and visualization techniques is required. 
Geographic maps, node-link diagrams, scientific visualizations, and multiple 2D visualizations as a large display are examples of potential immersive analytics visualizations~\cite{hurter2018fiberclay, yang2018origin, saenz2017reexamining, helbig2014concept, baltabayev2018virtual, chandler2015immersive}. In terms of interaction techniques, WIM, one-hand fly, two-hand fly, and Portal metaphors could be candidates for future research~\cite{laviola20173d}. A variation of Teleportation that supports smooth view transition is also a good candidate~\cite{heer2007animated, bhandari2018teleportation}.

Lastly, future research considering data scalability and distribution is required. The participants in this study were asked to interact with a hand-sized 3D plot containing 30 data points. Despite the fact that Kraus et al.~\cite{kraus2019impact} reported a negative effect of losing users' orientation within a big 3D visualization, a hand-size visualization might be unable to fully assist users in the visual analysis if the amount of data is large. Furthermore, we utilized a single dataset (Iris dataset) in order to ensure reasonable experiment durations and sub-sampled the attributes to avoid data points from forming any potential cluster. This random sub-sampling might introduce unexpected factors. 
To generalize our findings, further research with diverse data having different scalability and distribution is needed.
\section{Conclusion}

This work evaluates the three 3D user interaction techniques, Walking, Grab, and Teleportation, to explore data in IEs on their spatial working memory support. Our study design is motivated by the Corsi block-tapping task and spatial exploration to explore data in the immersive analytics setting. We compared the techniques in the VR and AR environments and reported the results. Our study results show that Walking supports spatial memory the most and Grab follows in both studies. Walking also has faster interaction time than Grab. We found that Teleportation supports spatial memory the least in the VR study, but many participants noted that it is the simplest way to explore VR environments. We also compared the Walking and Grab results in VR and AR and discussed the potential reasons explaining why they have different results. Our results showed Walking has no significant differences in distance errors, while Grab has a significant difference when longer movement is required. On the other hand, interaction times for both Walking and Grab were faster in the VR setting than in AR. 



\bibliographystyle{abbrv-doi}

\bibliography{0_ismar_main}
\end{document}